\documentclass[aps,showpacs,preprint,superscriptaddress,12pt]{revtex4}
\usepackage{graphicx}
\usepackage{bm}
\usepackage{txfonts}
\usepackage{subfigure}
\usepackage{epstopdf}

\begin{document}
\title{ Angular Distributions of Thomson Scattering in Combined Laser and Magnetic fields}
\author{Li Zhao}
\affiliation{College of Nuclear Science and Technology,  Beijing Normal University, Beijing 100875, China}
\affiliation{Key Laboratory of Beam Technology of the Ministry of Education, Beijing 100875, China}
\author{Zhijing Chen}
\affiliation{College of Nuclear Science and Technology,  Beijing Normal University, Beijing 100875, China}
\affiliation{Key Laboratory of Beam Technology of the Ministry of Education, Beijing 100875, China}
\author{Chun Jiang}
\affiliation{College of Nuclear Science and Technology,  Beijing Normal University, Beijing 100875, China}
\affiliation{Key Laboratory of Beam Technology of the Ministry of Education, Beijing 100875, China}
\author{Jian Huang}
\affiliation{College of Nuclear Science and Technology,  Beijing Normal University, Beijing 100875, China}
\affiliation{Key Laboratory of Beam Technology of the Ministry of Education, Beijing 100875, China}
\author{Chong Lv}
\affiliation{College of Nuclear Science and Technology,  Beijing Normal University, Beijing 100875, China}
\affiliation{Key Laboratory of Beam Technology of the Ministry of Education, Beijing 100875, China}
\author{Bai-Song Xie}
\affiliation{College of Nuclear Science and Technology,  Beijing Normal University, Beijing 100875, China}
\affiliation{Key Laboratory of Beam Technology of the Ministry of Education, Beijing 100875, China}
\affiliation{Beijing Radiation Center, Beijing 100875, China}
\author{ Hai-Bo Sang\footnote{Corresponding author. Email address: sanghb@bnu.edu.cn}}
\affiliation{College of Nuclear Science and Technology,  Beijing Normal University, Beijing 100875, China}
\affiliation{Key Laboratory of Beam Technology of the Ministry of Education, Beijing 100875, China}
\affiliation{Beijing Radiation Center, Beijing 100875, China}

\date{\today}
\begin{abstract}
Angular distributions of Thomson scattering are researched in the
combined fields with a circularly polarization of laser field and
a strong uniform magnetic field. The trajectories of the electron
and the dependence of it on the initial phase are also given. It
is found that the angular distributions with respect to the
azimuthal angle show twofold symmetry whatever the laser
intensity, the order of harmonics, the resonance parameter, and
the initial axial momentum are. On the other hand, the radiation
with respect to the polar angle is mainly distributed in two
regions which are roughly symmetric of the laser propagation. In
addition, the larger the laser intensity, the resonance
parameters, the initial axial momentum are, the closer radiation
is to the laser propagation direction. Besides, a new possibility
of X-ray production is indicated. That is to say, with appropriate
choice of laser and electron parameters, the high frequency part
of the Thomson scattering radiation can reach the frequency range
of X-ray($10^{17}Hz-10^{18}Hz$).

\end{abstract}

\maketitle

\section{Introduction}\label{introduction}
Due to the rapid development of ultra-strong ultra-short laser technology, a new stage in the study of the interaction between laser and matter is coming \cite{SCi-264-917,Natp-2-696,Nat-406-164,POP-8-1774,Natp-4-641,PRL-10-75}. Without doubt, Thomson scattering is a very important phenomenon during the interaction. Moreover, it is found that Thomson scattering has a potential to produced atto-second or even zepto-second X-ray pulses\cite{PRL-88-074801,PRE-72-066501}. And it is known that X-rays generated from Thomson scattering have many advantages. Thus, Thomson scattering which is produced by relativistic electrons in laser field has been put in progress extensively\cite{PRE-72-066501,PRE-48-3003,PRE-52-5425,PRE-54-2956,POP-9-4325,POP-10-2155,PRL-90-055002,PRA-58-3221,PRA-60-3276,PRA-60-2505,PRA-62-053809,PRA-61-043801,PRE-68-056501,PST-9-2,APL-95-161105,PRA-94-052102,EPL-117-44002}.

In the very beginning, Salamin \textit{et al.} research on a single electron in combined laser and magnetic field and obtain the general relativistic trajectory equations of the electron. It is also found that the forward spectrum consists of only two peaks: Thomson peak and magnetic peak\cite{PRA-58-3221,PRA-60-3276,PRA-60-2505}.
Then, the spectra of radiation in $\sin^{2}$ laser pulse with a uniform magnetic field is studied numerically, which shows that the forward and backward spectra are richer due to the uniform magnetic field\cite{PRA-60-2505}. Since then, extensive researches have been carried out to figure out some details related to Thomson scattering spectrum. For example, He \textit{et al.} find the phase dependence of the radiation spectra in a linearly polarized plane wave\cite{POP-9-4325}. Contemporarily, He \textit{et al.} reinvestigate similar problems about Thomson scattering in the combination of an intense laser field and a uniform magnetic field. The results show that the varying initial phases of the laser field bring about quite different effect for the circularly and linearly polarized laser field\cite{PRE-68-056501}. In particular, the spatial characteristics of Thomson scattering is firstly studied in a linearly polarized laser field by Lan \textit{et al.} in 2007\cite{PST-9-2}. It is shown that the spatial distributions of Thomson scattering depend sensitively on the initial phase and the symmetry of the spatial distributions only appears when the initial phase is 0 or $\pi$. In the last two years, the properties of Thomson back-scattering spectra are explored concretely in combined magnetic and laser fields \cite{PRA-94-052102,EPL-117-44002}, where the scale invariance and scaling law of the radiation spectrum with the relevant parameters such as the laser intensity, the initial axial momentum are found.

In this paper, in order to understand deeply the properties of Thomson scattering, the spatial angular distributions of the radiation spectrum are investigated, which is produced by moving electron in combined circularly polarized laser and magnetic fields. Based on the analytic trajectory of the electron, the spatial angular distribution with respect to the polar angle and the azimuthal angle through numerical integration is examined.

The paper is organized as follows. In Sec.II, the equations of the momentum and trajectory of the electron in combined laser and magnetic field are obtained. The computational formula of the emission power detected far away from the electron in a random direction is also given. In Sec.III, the trajectory of the electron and features of the angular distributions of Thomson scattering are presented. The main conclusions are given in the final section.

\section{basic equations}\label{basic}

In this paper, let's consider the spacial angular distributions of Thomson scattering that the electron (with mass $m$ and charge $-e$) moves in the simultaneous laser and magnetic field. Assuming that the external magnetic field of strength $B_{0}$, while the laser field is a circularly polarized plane wave with vector potential amplitude $A_{0}$ and frequency $\omega_{0}$. The laser propagation direction and the magnetic direction are both along the positive z-direction. By denoting the phase of the laser field as $\eta={\omega}_{0}t-\boldsymbol{k}\cdot\boldsymbol{r}$, where $\boldsymbol{k}$ and $\boldsymbol{r}$ are the laser wave vector and electron displacement vector respectively, the combined fields can be expressed by the total vector potential:

\begin{equation}
\label{eq1}\boldsymbol{A}=
{A}_{0}\left [-\sin {\eta}\hat{\textbf{i}}+\cos{\eta}\hat{\textbf{j}}\right ]+{B}_{0}x\hat{\textbf{j}},
\end{equation}

In order to simplify the calculations later, we shall normalized time by $1/\omega_{0}$, distance by $1/k_{0}$ and velocity by $c$, momentum by $mc$, vector potential by ${E}_{0}/m\omega_{0}c$, and the magnetic field by $e{B}_{0}/m\omega_{0}c$. Since an electron moves in a constant-amplitude circularly polarized laser filed, the constant of the motion is readily obtained as $\varsigma =\gamma-p_{z}=\gamma_{0}-p_{z0}$, where $\gamma$ is the electron relativistic factor, $p_{z}$ is the $z$ component of the electron relativistic momentum and $\gamma_{0},p_{z0}$ are the initial value of them. From Eq.(\ref{eq1}), the corresponding electric field $\boldsymbol{E}$ and magnetic field $\boldsymbol{B}$ can be obtained. Then, substituting them into the relativistic Lorentz force law, the momentum equations of the electron yields:

\begin{eqnarray}
\label{eq2}d^2 p_{x} /d \eta ^2+{\omega _{b}}^2 p_{x}= (\omega_{b} +1)a\sin{\eta},\\
\label{eq3}d^2 p_{y} /d \eta ^2+{\omega _{b}}^2 p_{y}=- (\omega_{b} +1)a\cos{\eta},
\end{eqnarray}
where $a=eA_{0}/mc^{2}$ is the normalized laser intensity and $\omega_{b}= b/ \varsigma$ is the cyclotron frequency of the electron motion in the combined field \emph{normalized by} $eB_{0}/m\omega_{0}c$.

Assuming that at $t=0$ the electron is static and located at $x_{0}=0,y_{0}=0,z_{0}=0$ with $\eta=\eta_{in}=-z_{0}$. By solving Eqs.(\ref{eq2}) and (\ref{eq3}) and then integrating, the momentum $\boldsymbol{p}(p_{x},p_{y},p_{z})$ and trajectory $\boldsymbol{r}(x,y,z)$ equations of the electron can be obtained:

\begin{equation}
\label{eq4}p_{x}=na\left \{ \sin{\eta}-\sin{\left[\omega_{b}\eta-\left( \omega_{b}-1\right)\eta_{in}\right]} \right \},
\end{equation}
\begin{equation}
\label{eq5}p_{y}=na\left \{ -\cos{\eta}+\cos{\left[\omega_{b}\eta-\left( \omega_{b}-1\right)\eta_{in}\right]} \right \},
\end{equation}
\begin{equation}
\label{eq6}p_{z}=\frac{2n^2a^2}{\varsigma}\sin^2{\left[\frac{\left(\omega_{b}-1 \right )\left(\eta-\eta_{in} \right )}{2}\right]}+\frac{1}{2\varsigma}-\frac{\varsigma}{2},
\end{equation}
\begin{equation}
\label{eq7}x(\eta)=\frac{ na}{\varsigma}\left \{ -\cos{\eta}+\frac{1}{\omega_{b}}\cos{\left[\omega_{b}\eta-\left ( \omega_{b}-1 \right )\eta_{in} \right ]} -\left ( \frac{1}{\omega_{b}}-1 \right )\cos{\eta_{in}}\right \},
\end{equation}
\begin{equation}
\label{eq8}y(\eta)=\frac{na}{\varsigma}\left \{ -\sin{\eta}+\frac{1}{\omega_{b}}\sin{\left[\omega_{b}\eta-\left ( \omega_{b}-1 \right )\eta_{in} \right ]} -\left ( \frac{1}{\omega_{b}}-1 \right )\sin{\eta_{in}}\right \},
\end{equation}
\begin{equation}
\label{eq9}z(\eta)=\left ( \frac{na}{\varsigma} \right )^{2}\left \{ \left ( \eta-\eta_{in} \right )-\frac{1}{\omega_{b}-1}\sin{\left [ \left ( \omega_{b}-1 \right )\left ( \eta-\eta_{in} \right ) \right ]} \right \}+\left ( \frac{1-\varsigma^2}{2\varsigma^2} \right )\left ( \eta-\eta_{in} \right ),
\end{equation}
where $n$ is the resonance parameter defined as $n=1/\left (\omega_{b}-1\right)$ in our previous work\cite{APL-95-161105}. Thus, the strength of external magnetic field $B_{0}=(1+1/n)(\sqrt{1+p_{z0}^{2}}-p_{z0})m\omega_{0}c/e\approx(1+1/n)(\sqrt{1+p_{z0}^{2}}-p_{z0})/\lambda(\mu m)$, which is related to the resonance parameter $n$ and the initial axial momentum $p_{z0}$ obviously. And from Eqs.(\ref{eq7}) - (\ref{eq9}), it can be seen that the electron trajectory is a rough helix and periodic.

The angular distributions of the emission power detected far away from the electron in the direction $\boldsymbol{n}$ (not to be confused with the magnetic resonance parameter $n$) can be calculated as Ref.\cite{Jakson}:
\begin{equation}
\label{eq10}\frac{d^2 I }{d\Omega d\omega }=\frac{e^2\omega^2}{4\pi ^2c}\left|\boldsymbol{n}\times \left[\boldsymbol{n}\times\boldsymbol{F}(\omega) \right ]\right|^2,
\end{equation}
\begin{equation}
\label{eq11}\boldsymbol{F}(\omega)=\frac{1}{\varsigma}\int_{-\infty }^{+\infty}d\eta\boldsymbol{p}(\eta)\exp{\left \{ i\omega\left [ \eta-\boldsymbol{n}\cdot\boldsymbol{r}(\eta)+z(\eta)\right ] \right \}},
\end{equation}
where the detected direction $\boldsymbol{n}=\sin\theta\cos\varphi\boldsymbol{x}+\sin\theta\sin\varphi\boldsymbol{y}+\cos\theta\boldsymbol{z}$.
Because the electron's motion is periodic with $T=2n\pi$, we simplify Eq.(\ref{eq11}) by an infinite series of delta functions and use the same computing method with our previous work\cite{PRA-94-052102} to calculate the emission power along any random direction. Similarly, there are a fundamental frequency $\omega_{1}=2\pi/(T-\boldsymbol{n} \cdot \boldsymbol{r}_{0}+z_{0})$ and harmonics of $\omega_{1}$ in this case too, so
\begin{equation}
\label{eq12}\boldsymbol{F}(\omega)=\frac{1}{\varsigma}\sum_{m=-\infty}^{+\infty}\boldsymbol{F}_{m}\delta\left (\omega-m\omega_{1} \right ),
\end{equation}
and the $m^{\rm{th}}$ amplitude is
\begin{equation}
\label{eq13}\boldsymbol{F}_{m}=\omega_{1}\int_{\eta_{in}}^{\eta_{in}+T}d\eta\boldsymbol{p}(\eta)\exp{\left \{ im2\pi h(\eta) \right \}},
\end{equation}
with
\begin{equation}
\label{eq14}h(\eta)=\frac{\eta-\boldsymbol{n}\cdot\boldsymbol{r}(\eta)+z(\eta)}{T-\boldsymbol{n}\cdot\boldsymbol{r}_{0}+z_{0}},
\end{equation}
where $\boldsymbol{r}_{0}=(0,0,T\left [\left ( \frac{na}{\varsigma} \right )^2+\frac{1}{2}\left ( \frac{1}{\varsigma^2}-1 \right )  \right ])$ is the drift displacement vector of the electron during one period.

Thus, based on Eq.(\ref{eq4})-Eq.(\ref{eq14}), the angular distributions of Thomson scattering when an electron moves in the laser and magnetic fields can be calculated numerically. For simplicity, in the paper, the parameters are chosen to make the radiation reaction effect(RRE) ignorable, which is as same as our previous work\cite{PRA-94-052102}.

\section{numerical results and analysis}\label{numeric}

In the following, the spatial features of the Thomson scattering are researched from two perspectives. One is spatial distributions with respect to the azimuthal angle $\varphi$, where the polar angle $\theta=\pi/2$. Accordingly, the detected direction $\boldsymbol{n}=(\cos\varphi,\sin\varphi,0)$, the fundamental frequency $\omega_{1}={2\pi}/({2\pi n+z_{0}})$, and $h(\eta)=[{\eta-\cos\varphi x(\eta)-\sin\varphi y(\eta)}+z(\eta)]/[{2\pi n+z_{0}}]$. Another one is spatial distributions with respect to the polar angle $\theta$, where the azimuthal angle $\varphi=0$. Accordingly, the detected direction $\boldsymbol{n}=(\sin\theta,0,\cos\theta)$, the fundamental frequency $\omega_{1}={2\pi}/[{2\pi n+(1-\cos\theta)z_{0}}]$, and $h(\eta)=[{\eta-\sin\theta x(\eta)+(1-\cos\theta)z(\eta)}]/[{2\pi n+(1-\cos\theta)z_{0}}]$.

\subsection{The electron's trajectories }

\begin{figure}[htbp]\suppressfloats
\includegraphics[width=15cm]{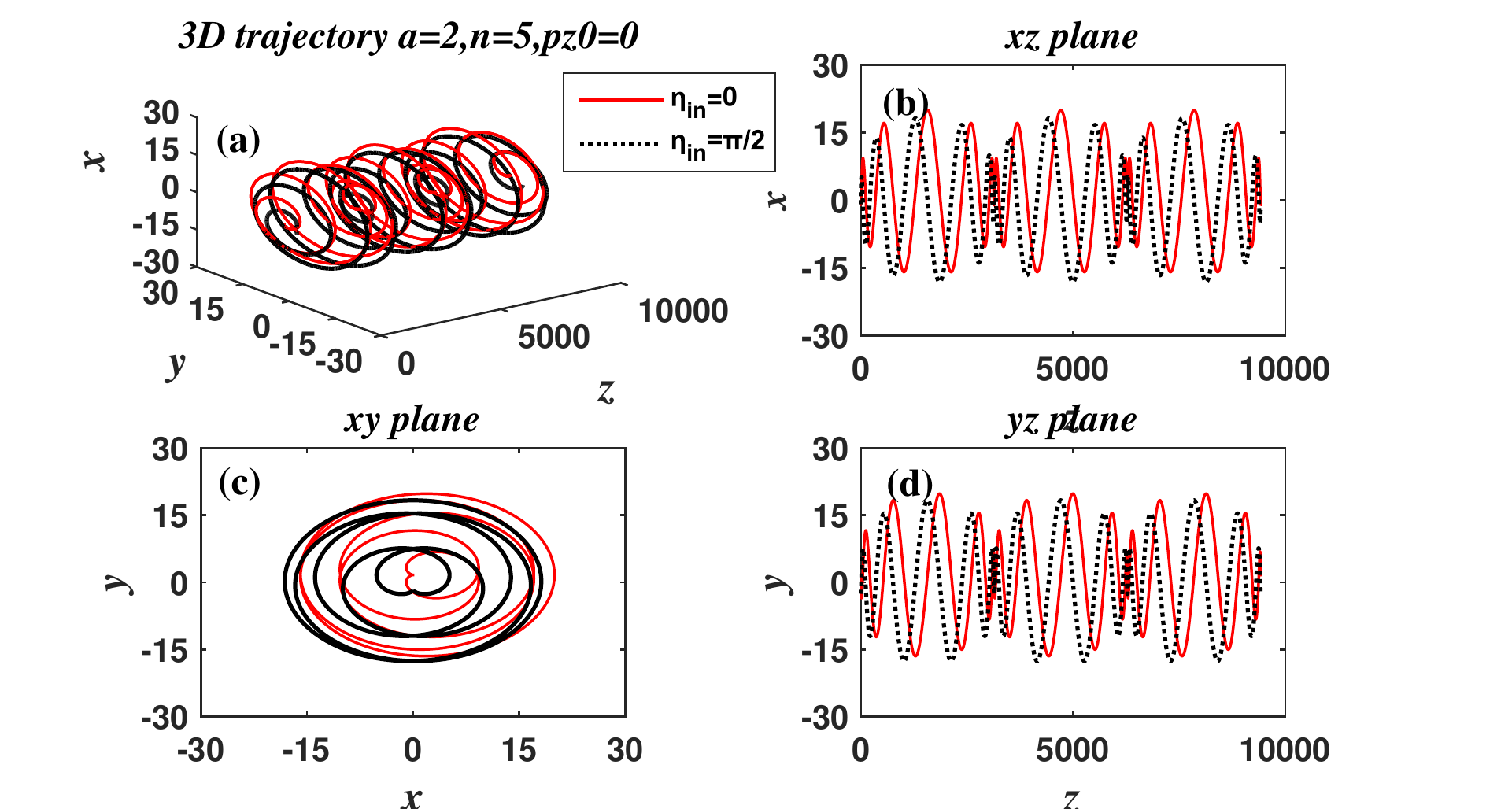}
\caption{\label{Fig1}(color online)Electron's trajectories in the combined laser and magnetic fields when $\eta_{in}=0,\pi/2$. The initial axial momentum $p_{z0}=0$, the laser intensity $a=2$, and the resonance parameter $n=5$. The phase $\eta$ varies from 0 to $30\pi$.}.
\end{figure}

First of all, let's study the electron's trajectories in the combined laser and magnetic fields. As an example, the electron's trajectories for various $\eta_{in}$ are shown in Fig.\ref{Fig1}, where the initial axial momentum $p_{z0}$, the laser intensity $a$, and the resonance parameter $n$ are set as $p_{z0}=0,a=2,n=5$ respectively. Due to the periodicity of phase $\eta$, the electron's trajectories are calculated over three optical cycles(the phase $\eta$ varies from 0 to $30\pi$). Notably, the following results have nothing to do with the optical cycles. The three dimensions trajectories of the electron are shown in Fig.\ref{Fig1}(a). Obviously, the electron experiences a helically periodic motion in the combined fields. As shown in Fig.\ref{Fig1}(b),(d), the projections of the electron's trajectory in the $xz$ plane as well as the $yz$ one are oscillating. It presents that the electron orbits drift in $z$ direction. And it can be seen that over one period $\textit{T}$, the electron undergoes a net displacement $\boldsymbol{r}_{0}=(0,0,z_{0})$, which is consistent with our theory mentioned above. By comparison, it is found that the displacements are same both at $\eta_{in}=0$ and $\eta_{in}=\pi/2$.  Then, when we focus on the $xoy$ plane shown in Fig.\ref{Fig1}(c), it can be seen that the electron moves like drawing circles around and the shapes of the trajectories remain unchanged for different initial phase. And from Fig.\ref{Fig1}(c), it is found that when the initial phase of the laser $\eta_{in}$ is different, the moving direction of the electron from the initial position is different.

\subsection{The spatial distribution of the Thomson scattering }
\begin{figure}[htbp]\suppressfloats
\includegraphics[width=7.5cm]{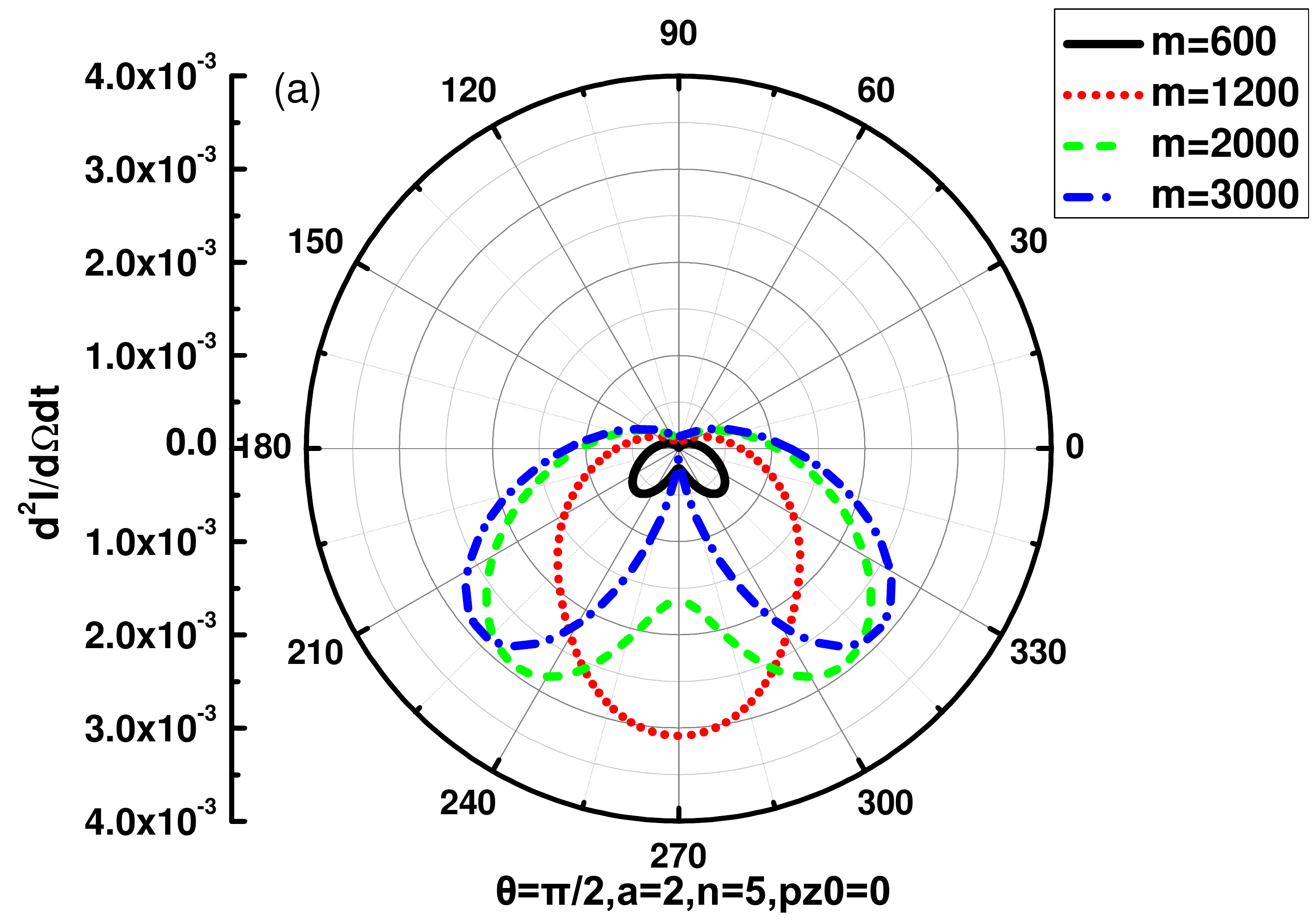}
\includegraphics[width=7.5cm]{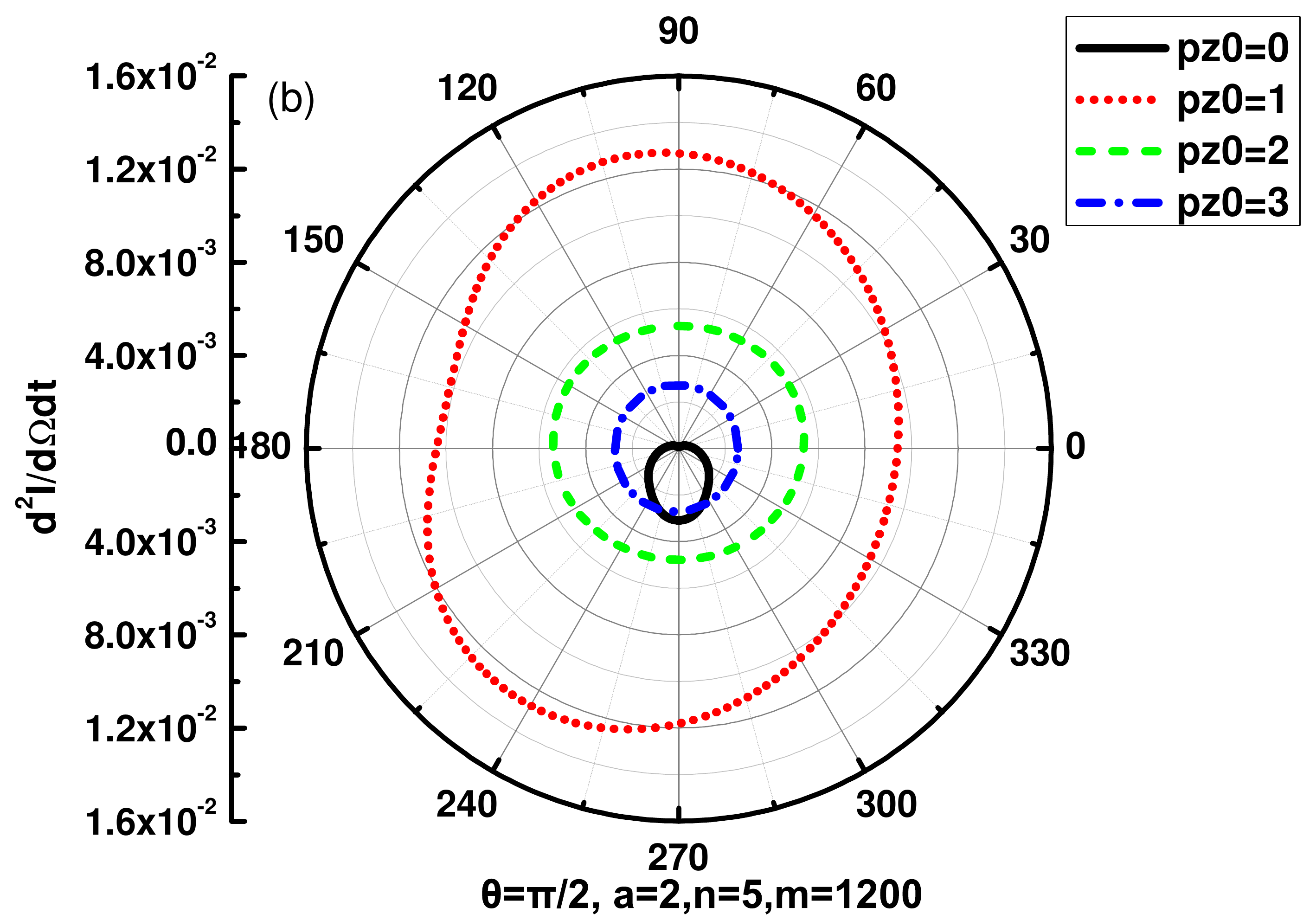}
\includegraphics[width=7.5cm]{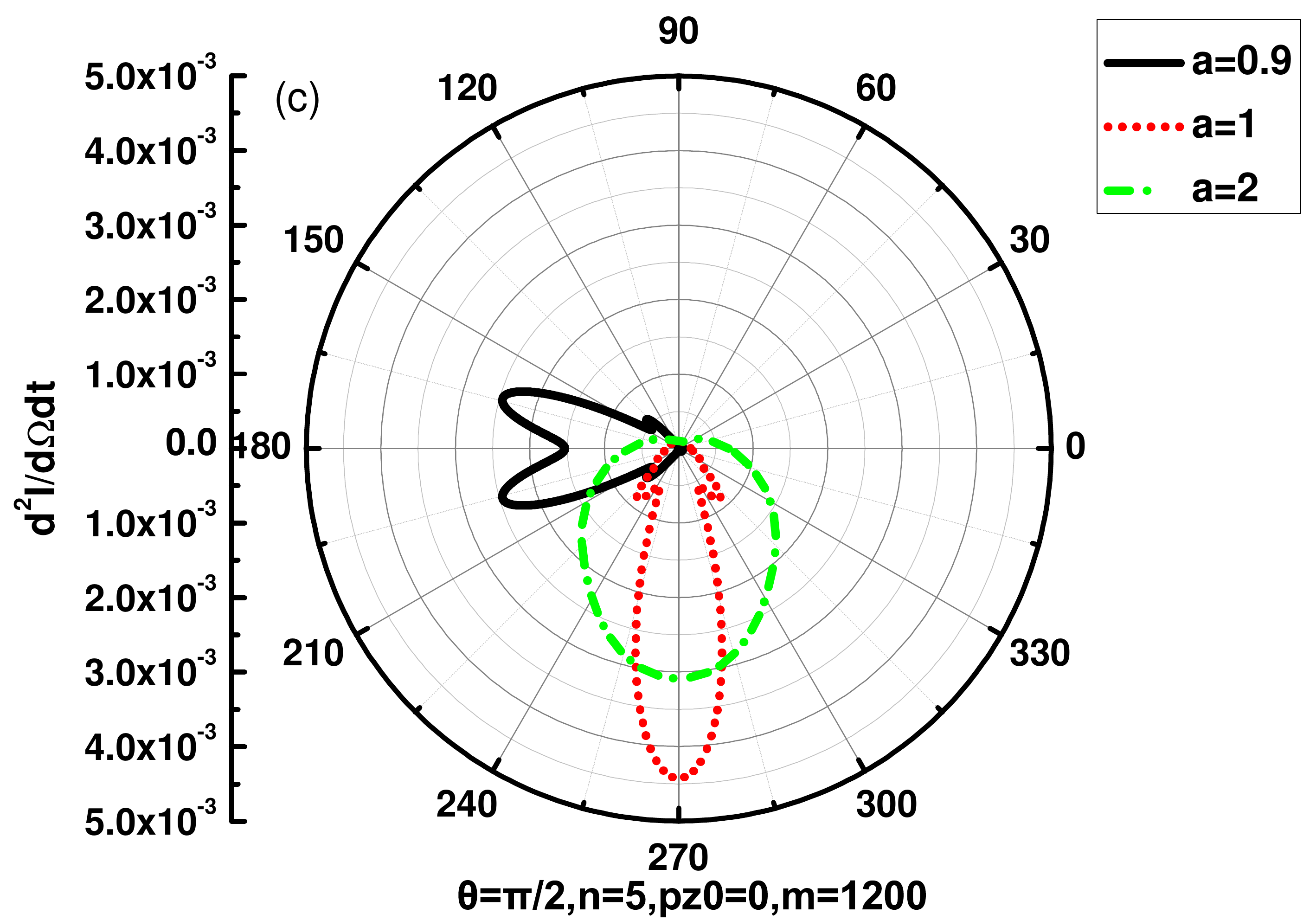}
\includegraphics[width=7.5cm]{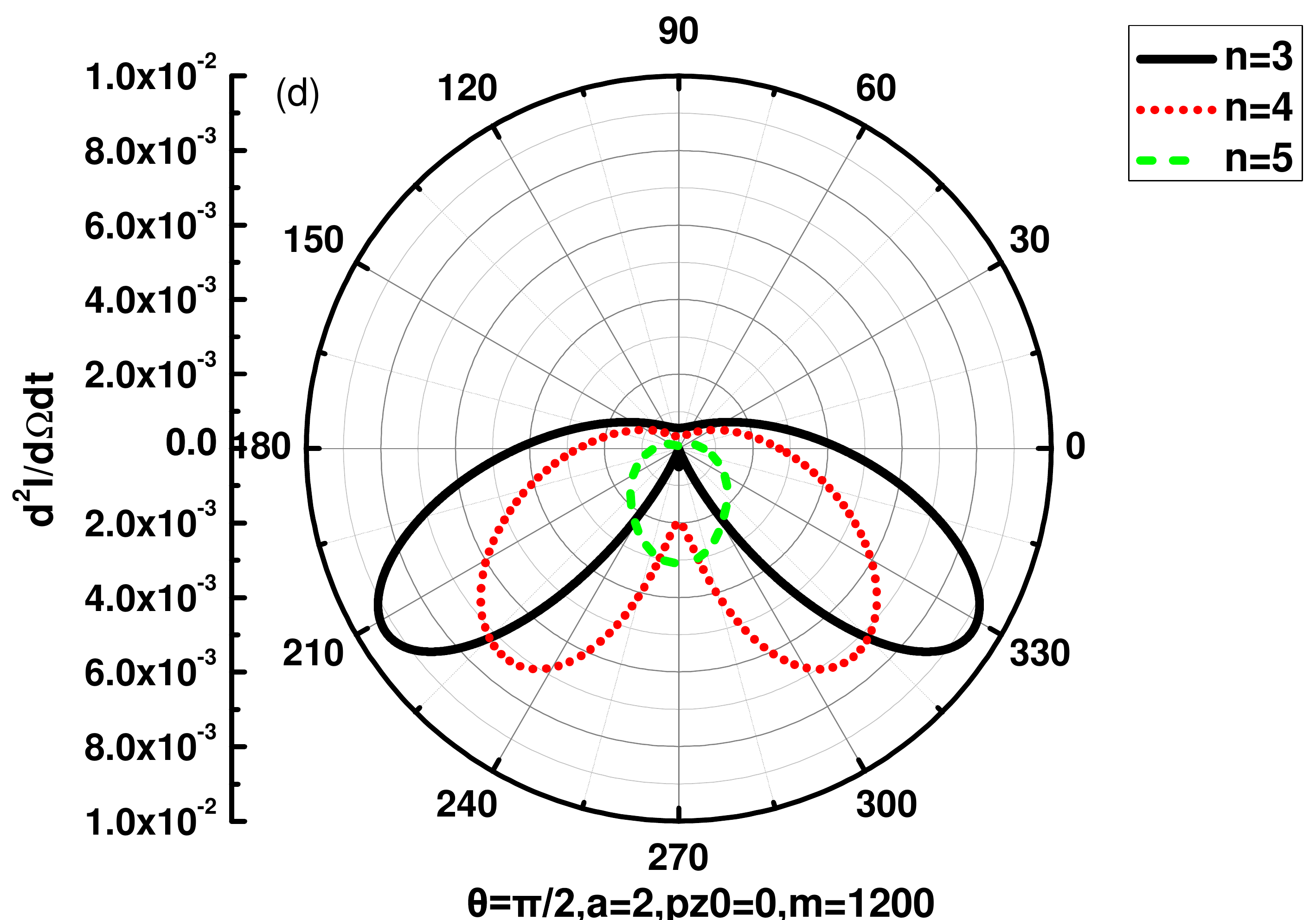}
\caption{\label{Fig2}(color online)Spatial distributions of the $m^{\rm{th}}$ harmonic radiation with respect to the azimuthal angle $\varphi$ (normalized by $\textit{e}^{2}/4\pi^{2}\textit{c}$ ). The different $m$, $p_{z0}$, $a$ and $n$ are shown in (a), (b), (c) and (d) respectively. The wavelength $\lambda=1\mu\textit{m}$, and the polar angle $\theta=\pi/2$. (a) $a=2$, $n=5$, $p_{z0}=0$,$m=600,1200,2000,3000$; (b)$a=2$,  $n=5$, $m=1200$, $p_{z0}=0,1,2,3$; (c) $n=5$, $p_{z0}=0$, $m=1200$, $a=0.9,1,2$; (d) $a=2$, $p_{z0}=0$, $m=1200$, $n=3,4,5$.}.
\end{figure}

The above results show that the electrons, injected into the laser field at different phases, undergo roughly the same movements, when other parameters are fixed. Thus for convenience, in the following simulations the initial phase is set as $\eta_{in}=0$.

Next, let's concentrate on the spatial characteristics of Thomson scattering in the combined laser and magnetic fields. First of all, the angular distributions of the emission with respect to the azimuthal angle $\varphi$ are investigated. From our previous work Ref.\cite{PRA-94-052102}, it is known that the Thomson scattering spectrum is consisted of a series of harmonics. Here, $m$ represents the order of the harmonics. The azimuthal angle distributions of the emitted power for various parameters with the harmonics order $m$, the laser intensity $a$, the resonance parameter $n$, and the axial initial momentum $p_{z0}$ are shown in Fig.\ref{Fig2}. The relevant parameters are chosen as (a) $a=2$, $n=5$, $p_{z0}=0$, $m=600,1200,2000,3000$; (b)$a=2$, $n=5$, $m=1200$, $p_{z0}=0,1,2,3$; (c) $n=5$, $p_{z0}=0$, $m=1200$, $a=0.9,1,2$; (d) $a=2$, $p_{z0}=0$, $m=1200$, $n=3,4,5$. We also suppose that the wavelength of laser is $1\mu\textit{m}$ and the polar angle $\theta=\pi/2$. Firstly, the spatial distributions of radiation with respect to the azimuthal angle $\varphi$ for different orders of harmonic are shown in Fig.\ref{Fig2}(a). It can be seen that for different harmonic orders the radiation distributions exhibit different shapes. But, there is one thing in common, that is, they are all symmetric with respect to $y$ axis. Also, it is found that the strongest radiation mainly distributes between $\varphi=\pi$ and $\varphi=2\pi$. Secondly, let's focus on Fig.\ref{Fig2}(b), where the angular distributions of the emission with respect to the azimuthal angle $\varphi$ for different initial axial momentum $p_{z0}$ are shown. Symmetry of radiation distribution still exists, but it is no longer symmetric with respect to $y$ axis when the initial axial momentum $p_{z0}=1$. Besides, when the initial axial momentum $p_{z0}\neq0$, the strongest radiation is approximately uniformly distributed. That is to say, whether the electron is stationary initially has an effect on the distribution of Thomson scattering. Thirdly, as shown in Fig.\ref{Fig2}(c), the symmetry of radiation distributions does not be broken if changing the laser intensity while other parameters are fixed. Specially, when the laser intensity $a=0.9$, the angular radiation distribution is symmetric with respect to $x$ axis not $y$ axis. Lastly, the properties of the spatial angular distributions for various resonance parameters are presented in Fig.\ref{Fig2}(d). It can be seen that the radiation distribution is still symmetric with respect to $y$ axis. And, in this case, the strongest radiation mainly distributes between $\varphi=\pi$ and $\varphi=2\pi$.

In general, the angular radiation distributions with respect to the azimuthal angle $\varphi$ show twofold symmetry. When the electron is static initially, the strongest radiation is mainly concentrated in a certain angle range. On the contrary, the radiation is approximately uniformly distributed with respect to the azimuthal angle $\varphi$.

\begin{figure}[htbp]\suppressfloats
\includegraphics[width=7.5cm]{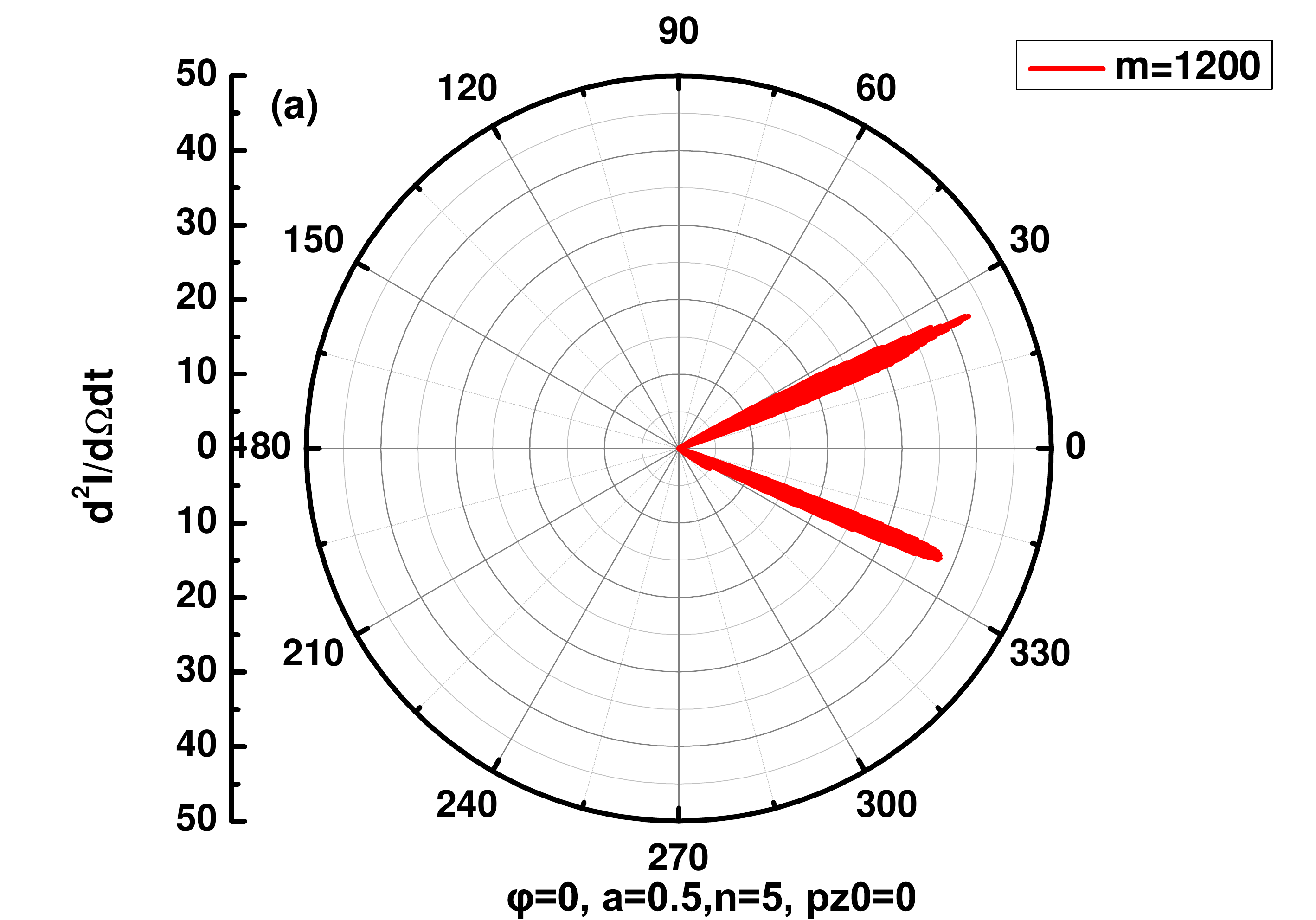}
\includegraphics[width=7.5cm]{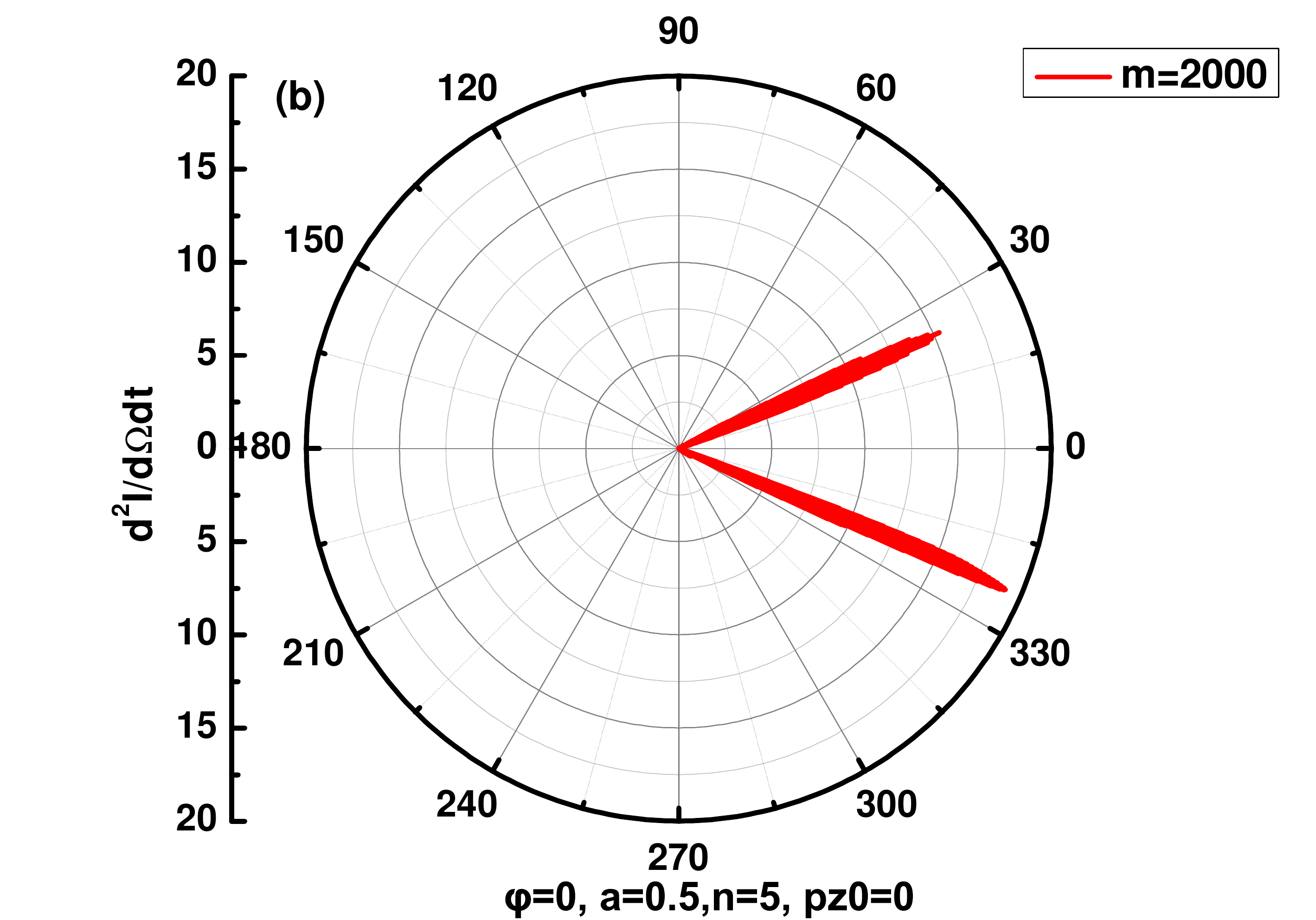}
\includegraphics[width=7.5cm]{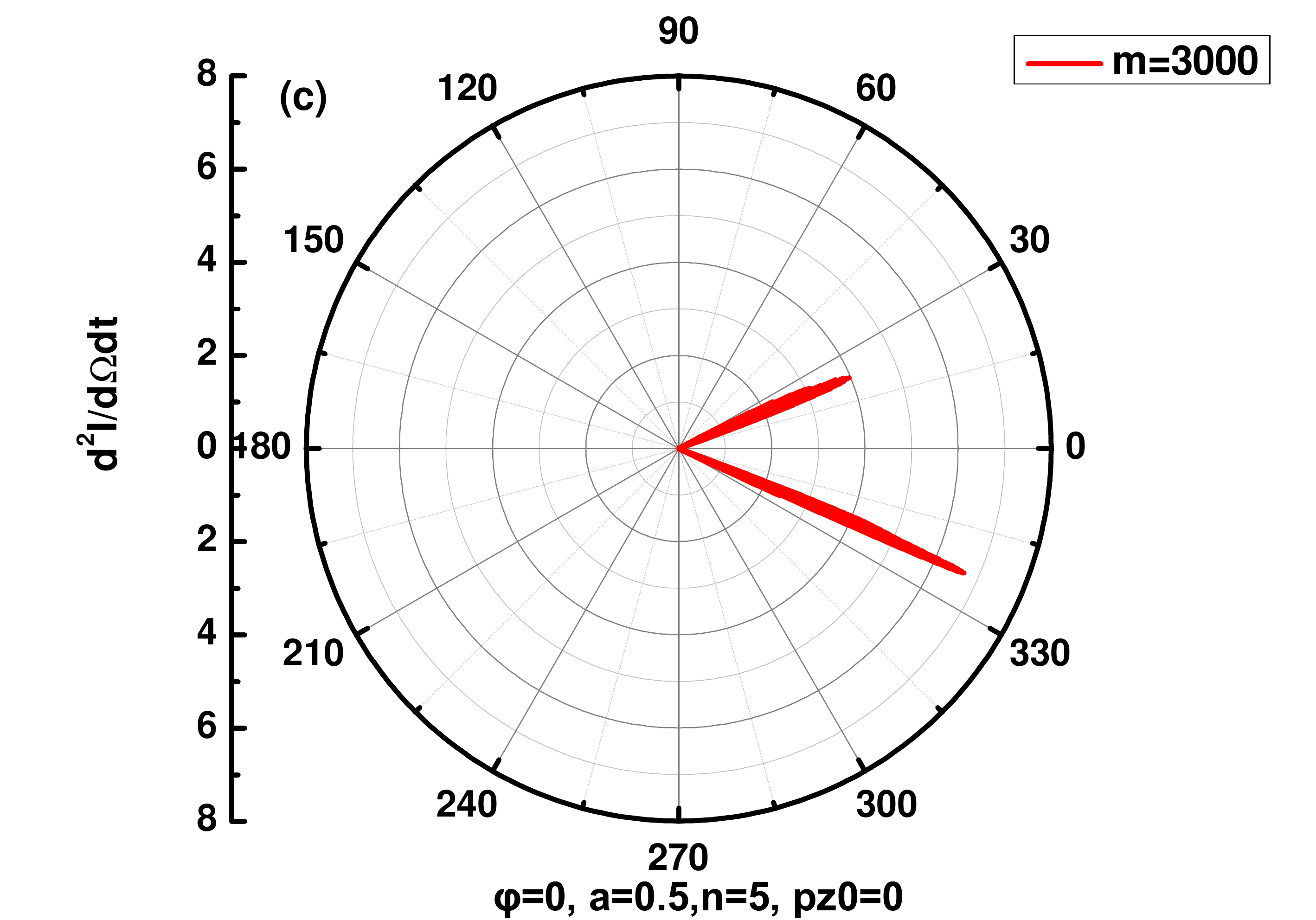}
\includegraphics[width=7.5cm]{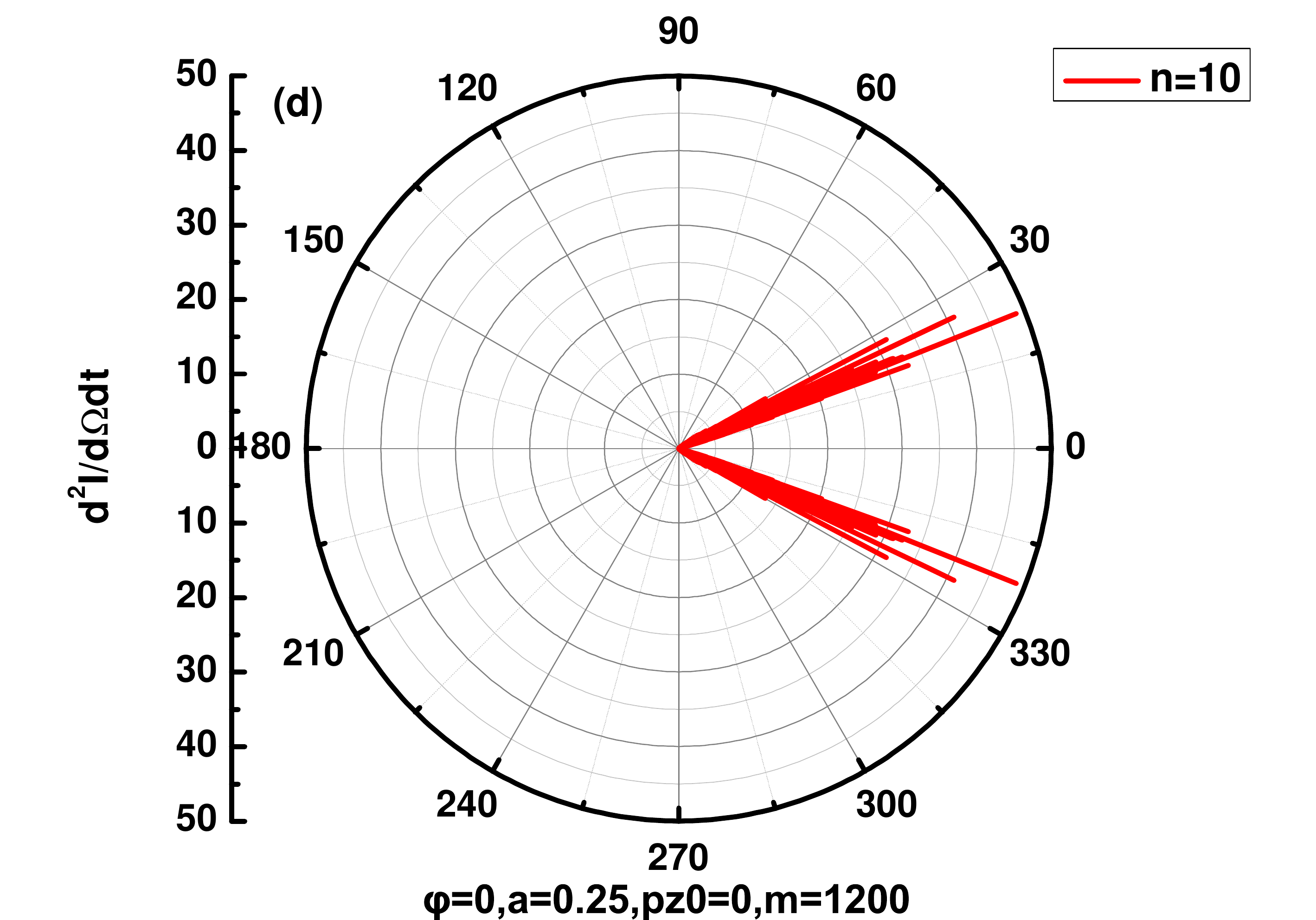}
\includegraphics[width=7.5cm]{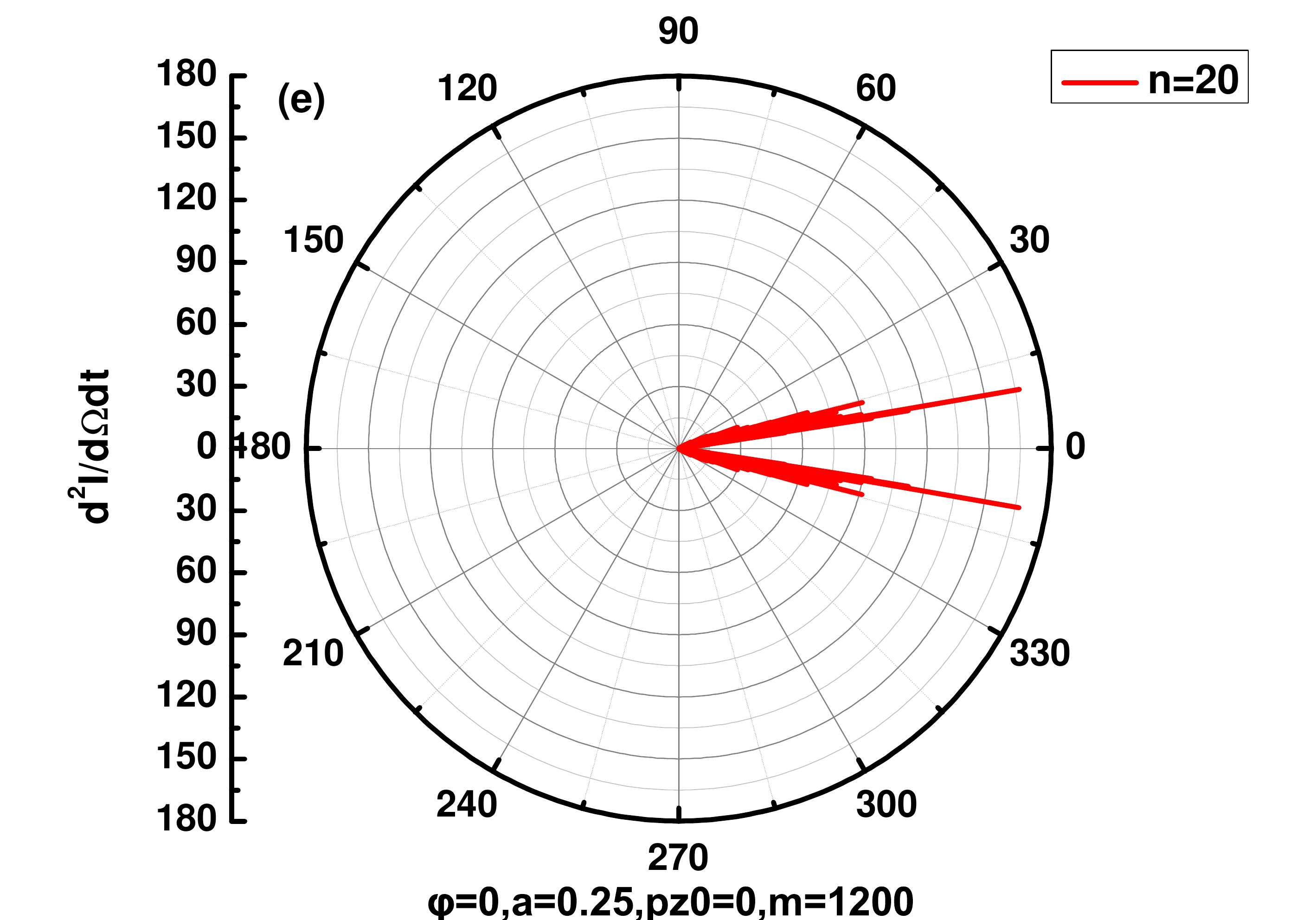}
\includegraphics[width=7.5cm]{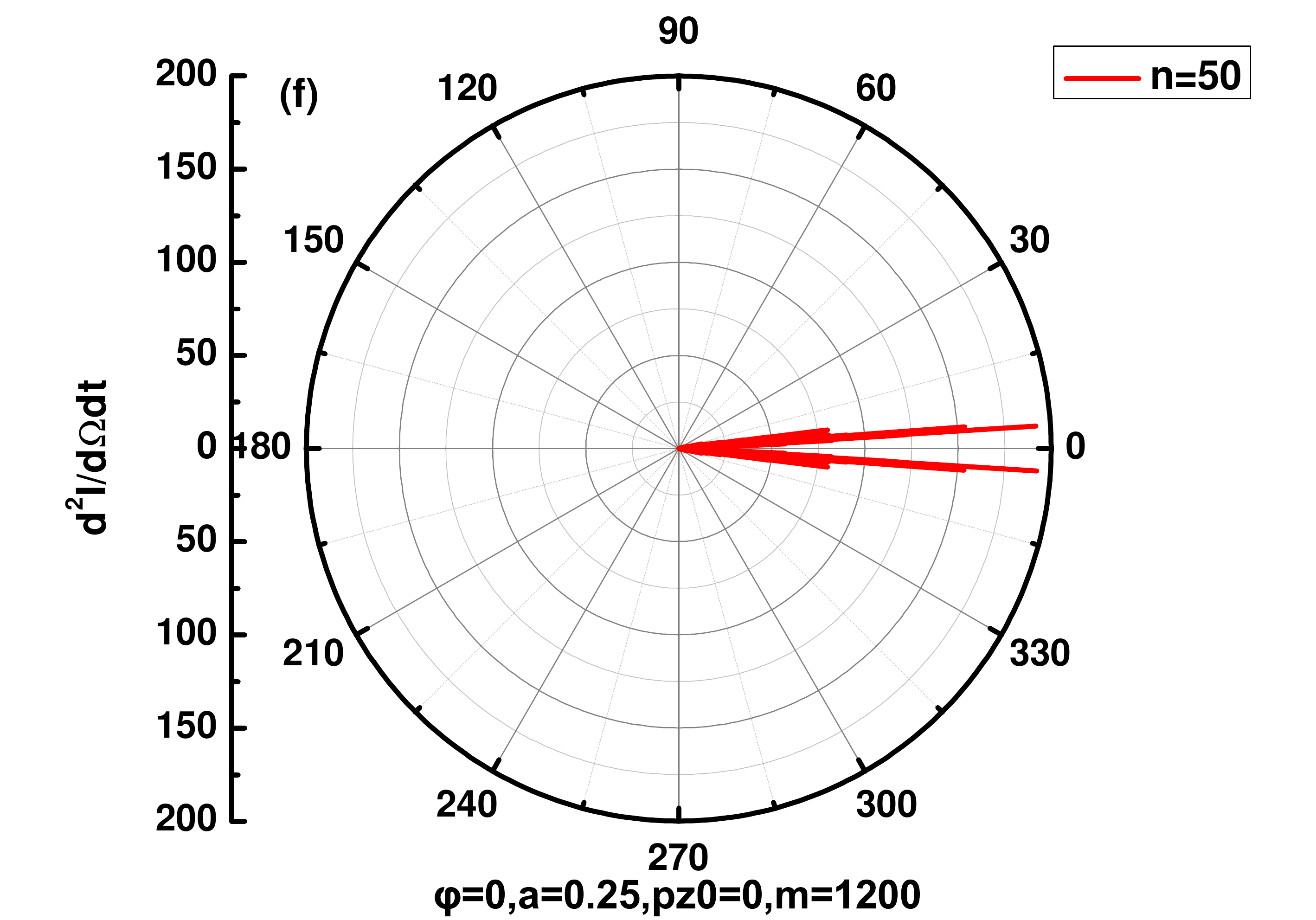}
\caption{\label{Fig3}(color online)Spatial distributions of the $m^{\rm{th}}$ harmonic radiation with respect to the polar angle $\theta$ (normalized by $\textit{e}^{2}/4\pi^{2}\textit{c}$ ). The wavelength $\lambda=1\mu\textit{m}$ and the azimuthal angle $\varphi=0$. (a)-(c): $a=0.5$, $n=5$, $p_{z0}=0$, $m=1200,2000,3000$; (d)-(f): $a=2$, $p_{z0}=0$, $m=1200$, $n=10,20,50$.}.
\end{figure}

On the other hand, the angular distributions of the emission in the combined laser and magnetic fields with respect to the polar angle $\theta$ are investigated. The results are shown in Fig.\ref{Fig3} for different orders of harmonic radiation $m$ and resonance parameter $n$, as (a)-(c) $a=0.5$, $n=5$, $p_{z0}=0$, $m=1200,2000,3000$ and (d)-(f) $a=2$, $p_{z0}=0$, $m=1200$, $n=10,20,50$. It is apparently that the spatial distribution of the strongest radiation with respect to the polar angle consists of two lobes collimated in the forward direction. One is in the $+xz$ plane, the other is in the $xz$ plane. From Fig.\ref{Fig3}(a),(b),(c), one can see that the maximum of the radiation energy emitted per unit solid angle $d\Omega$ and per unit frequency interval $d\omega$ decreases when the order of the harmonics increases from 1200 to 3000. In addition, the maximum radiation strength appears at the same polar angle even for different harmonic orders, which are axial symmetry of the laser propagation direction, although the distributions in the upper plane and the lower one are not exactly the same. At the same time, as shown in Fig.\ref{Fig3}(d)-(f), with the increasing of resonance parameter $n$, the emission strength increases too, while the angle between the two regions gets smaller. That is to say, the larger the resonance parameter $n$ is, the closer the radiation distribution is to the laser propagation direction. Moreover, under this condition, the radiation distributions are symmetric well with respect to the laser-propagation direction(the $z$ axis). By the way, based on the conclusions above, in the following study we will focus on the spatial distributions of the $1200^{th}$ harmonic radiation.

\begin{figure}[htbp]\suppressfloats
\includegraphics[width=7.5cm]{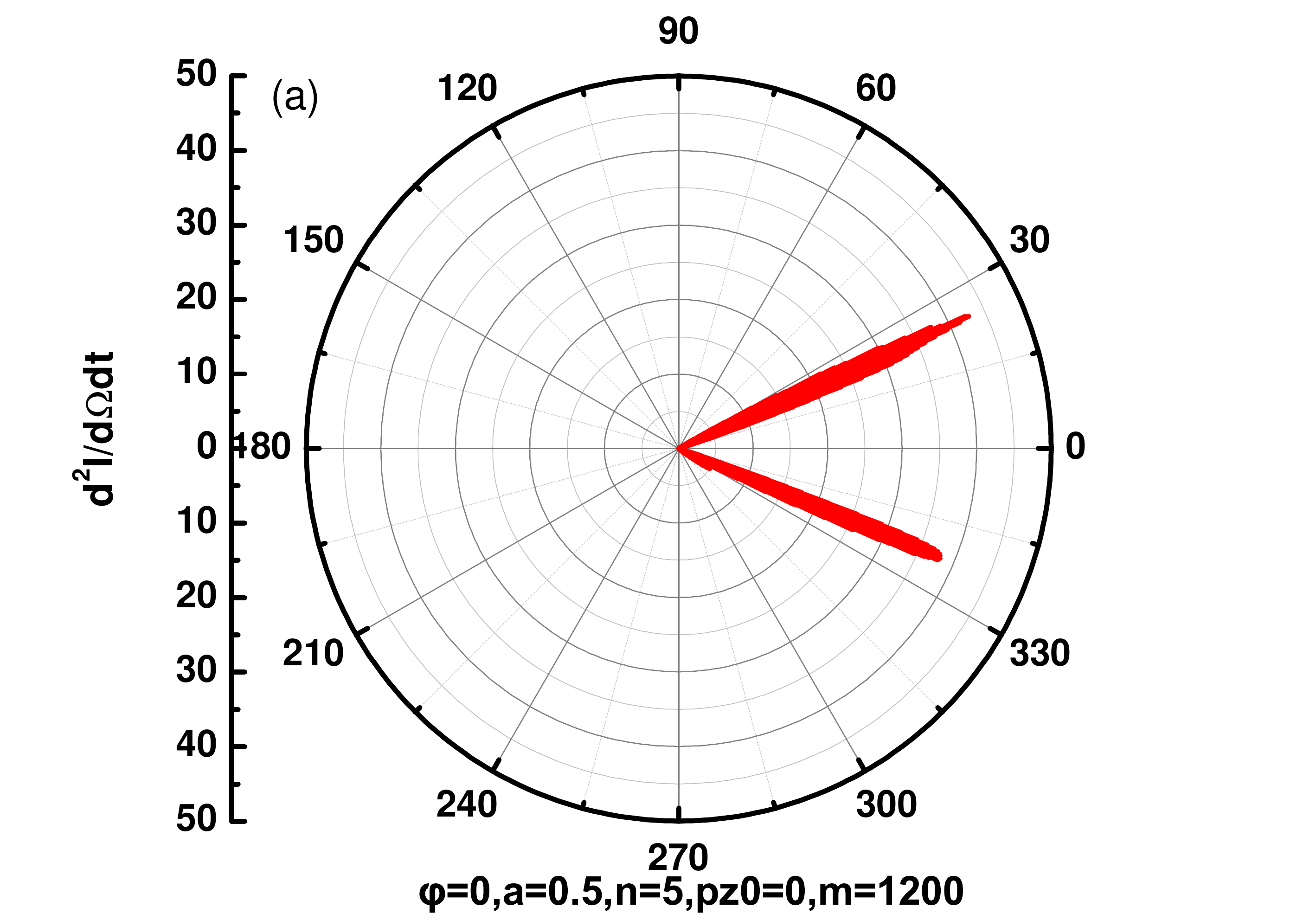}
\includegraphics[width=7.5cm]{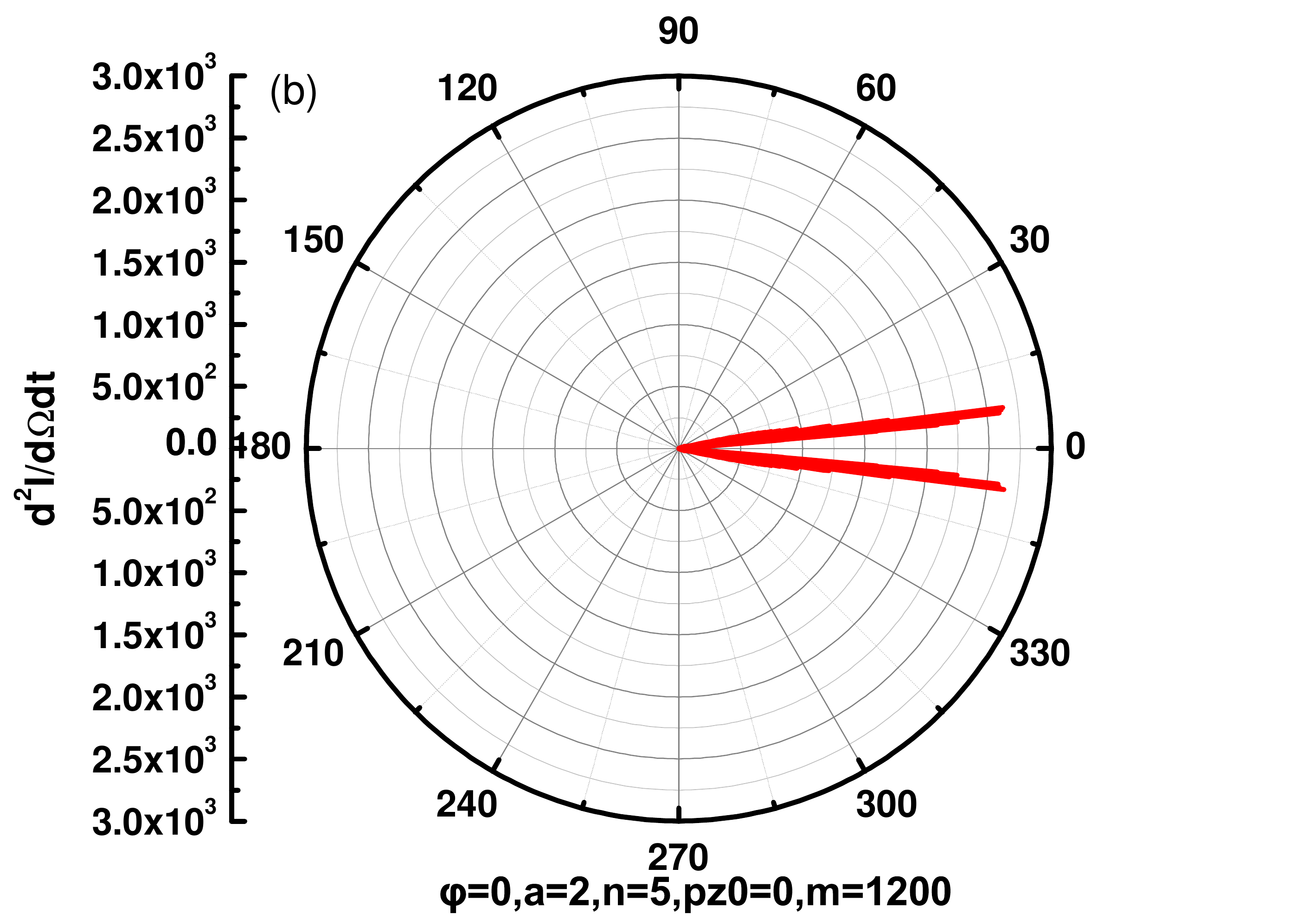}
\includegraphics[width=7.5cm]{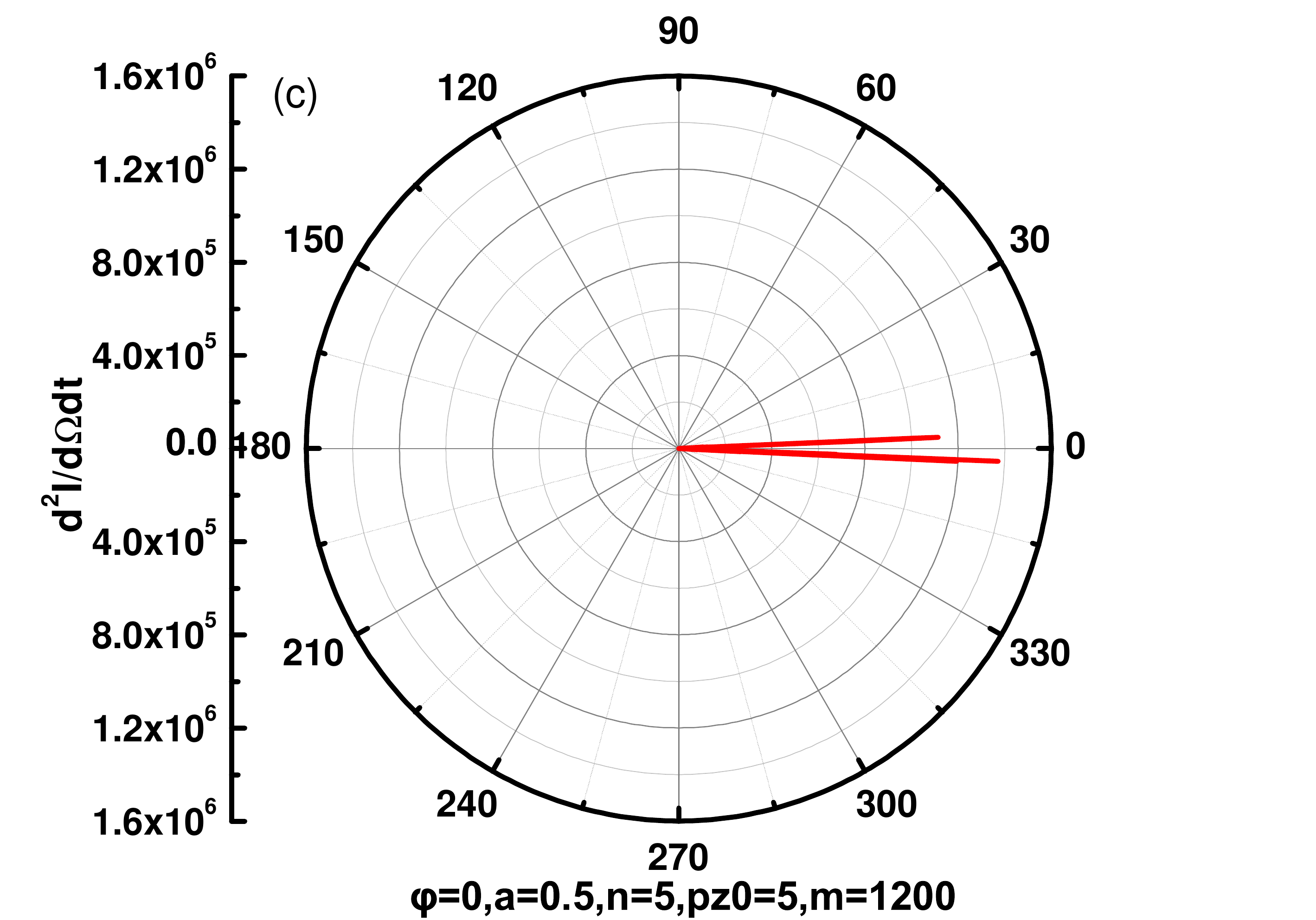}
\includegraphics[width=7.5cm]{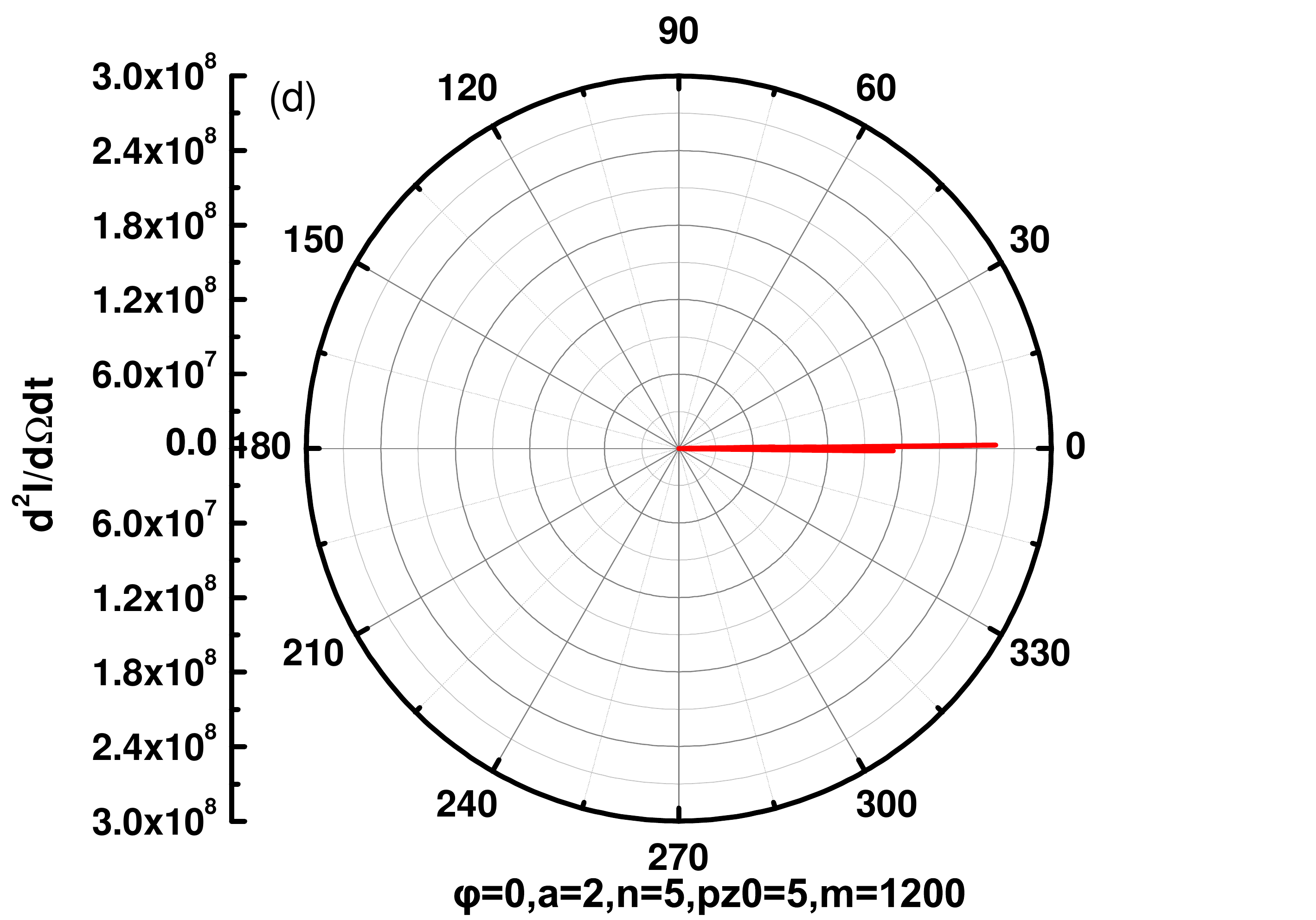}
\caption{\label{Fig4}(color online)Spatial distributions of the $1200^{\rm{th}}$ harmonic radiation with respect to the polar angle $\theta$ (normalized by $\textit{e}^{2}/4\pi^{2}\textit{c}$ ). The wavelength $\lambda=1\mu\textit{m}$,  the resonance parameter $n=5$ and the azimuthal angle $\varphi=0$. (a) $a=0.5$, $p_{z0}=0$; (b) $a=2$, $p_{z0}=0$; (c) $a=0.5$, $p_{z0}=5$; (d) $a=2$, $p_{z0}=5$.}.
\end{figure}

Meanwhile, the dependence of the spatial distribution of the radiation on the laser intensity and the axial initial momentum of the electron is also researched. In Fig.\ref{Fig4}, the spatial distributions of the $1200^{\rm{th}}$ harmonic radiation are plotted for two sets of different parameters. One set in (a) - (b) is for $n=5$, $p_{z0}=0$ but with different laser intensity $a=0.5,2$, and the other set in (c)-(d) is the same as (a)-(b) except that the initial axial momentum is changed to $p_{z0}=5$. By comparing the two charts on the left with the other two on the right, it is firstly found that when the laser intensity $a$ increases from 0.5 to 2 the angle between the two regions gets smaller, and the strength of the radiation increases sharply by two or three orders of magnitude. Secondly, with the initial axial momentum $p_{z0}$ becoming larger, the two radiation distribution regions  get closer to the laser-propagation direction, and the radiation gets stronger too. Specially, as shown in Fig.\ref{Fig4}(d), the radiation is mainly in the laser-propagation direction($+z$). Namely, the Thomson forward scattering dominates under this case.

To sum up, the radiation with respect to the polar angle $\theta$ is mainly distributed in two regions and this two regions are roughly symmetric of the laser propagation direction. Moreover, in case of the ignorance of RRE, the strength of radiation increases as the laser intensity $a$, the initial axial momentum $p_{z0}$ and the resonance parameters $n$ increase. It is possible that the radiation can be concentrated in the forward direction if an appropriate set of parameters is chosen.

\subsection{The optimum observation direction }

\begin{figure}[htbp]\suppressfloats
\includegraphics[width=7.5cm]{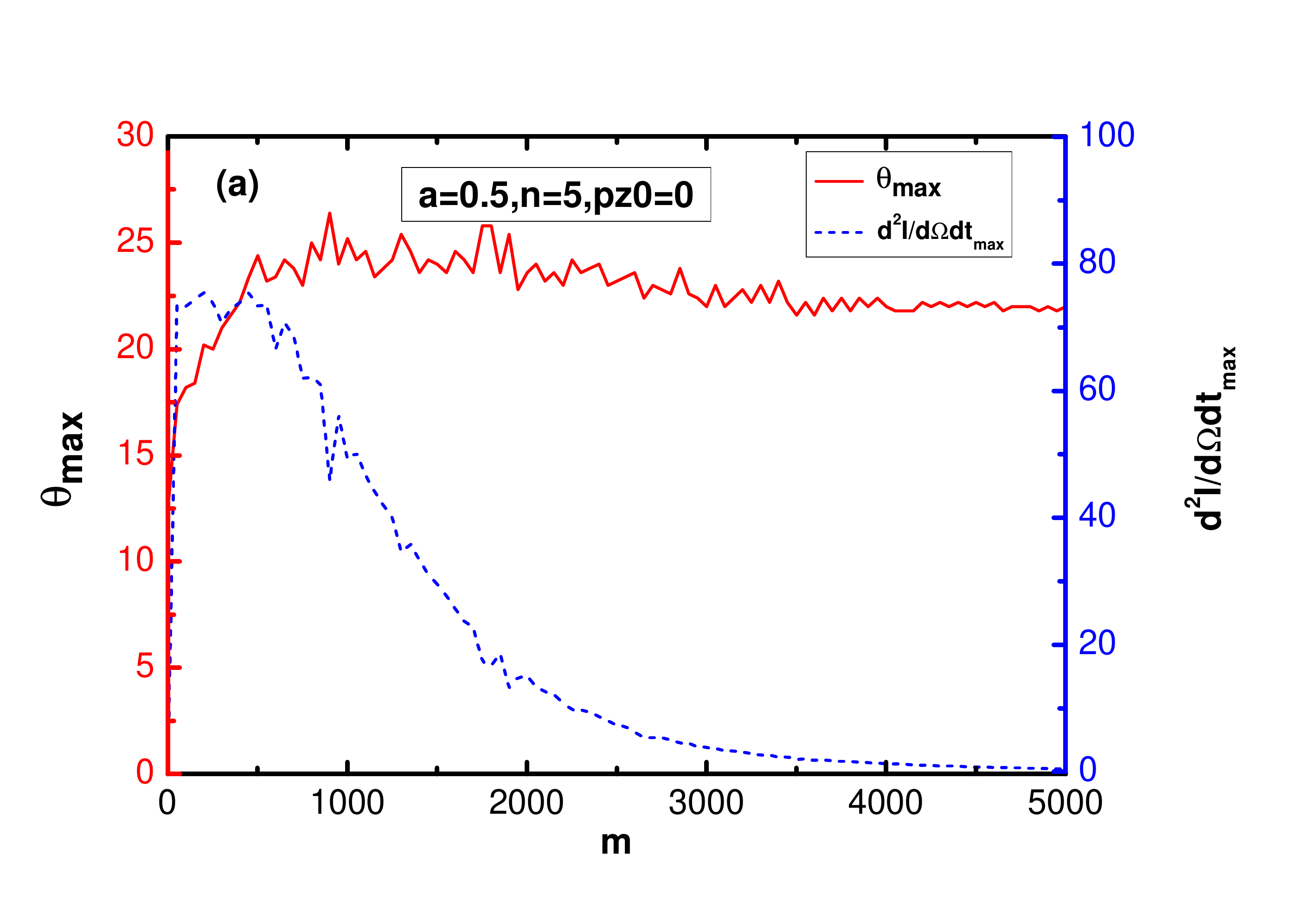}
\includegraphics[width=7.5cm]{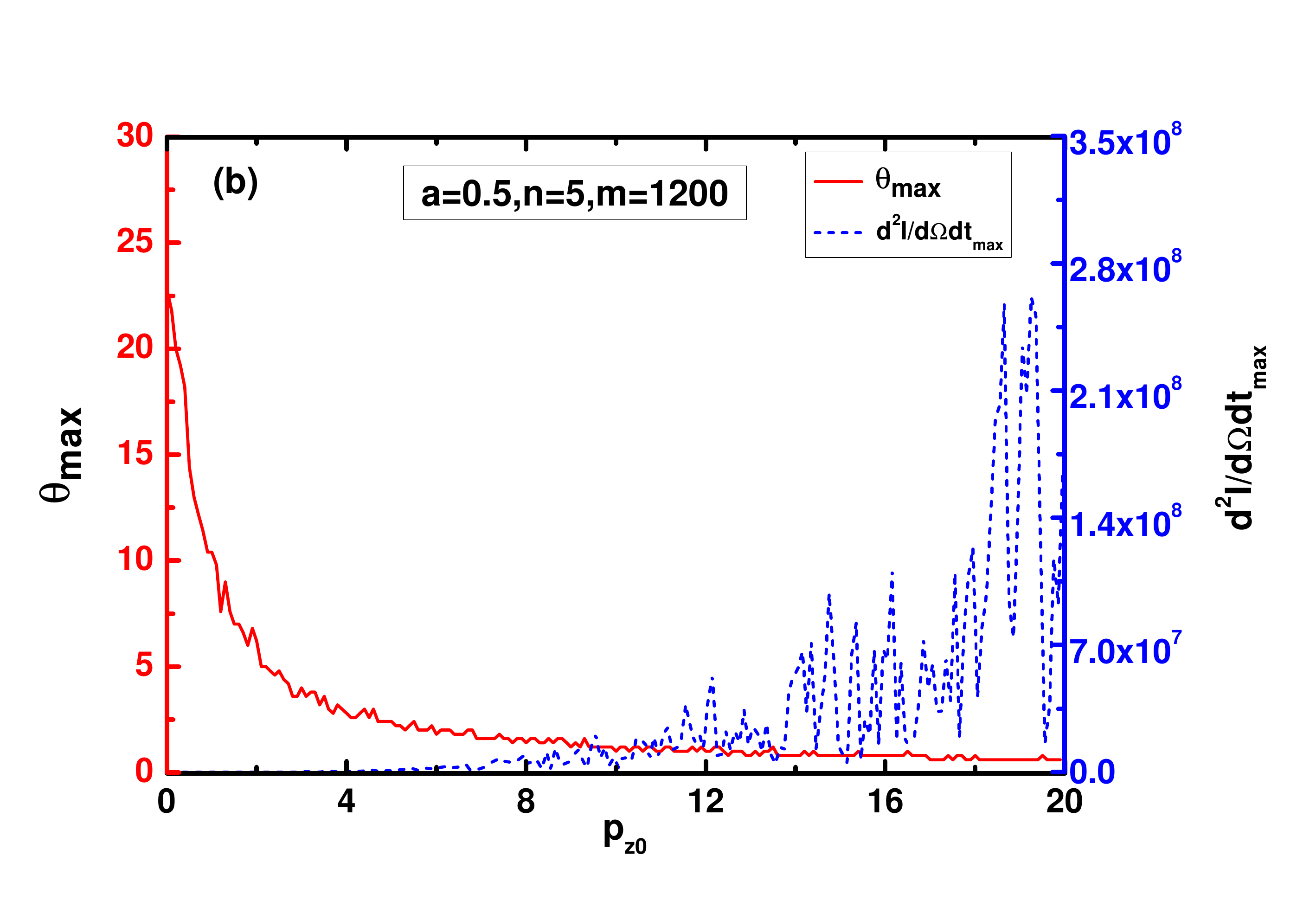}
\includegraphics[width=7.5cm]{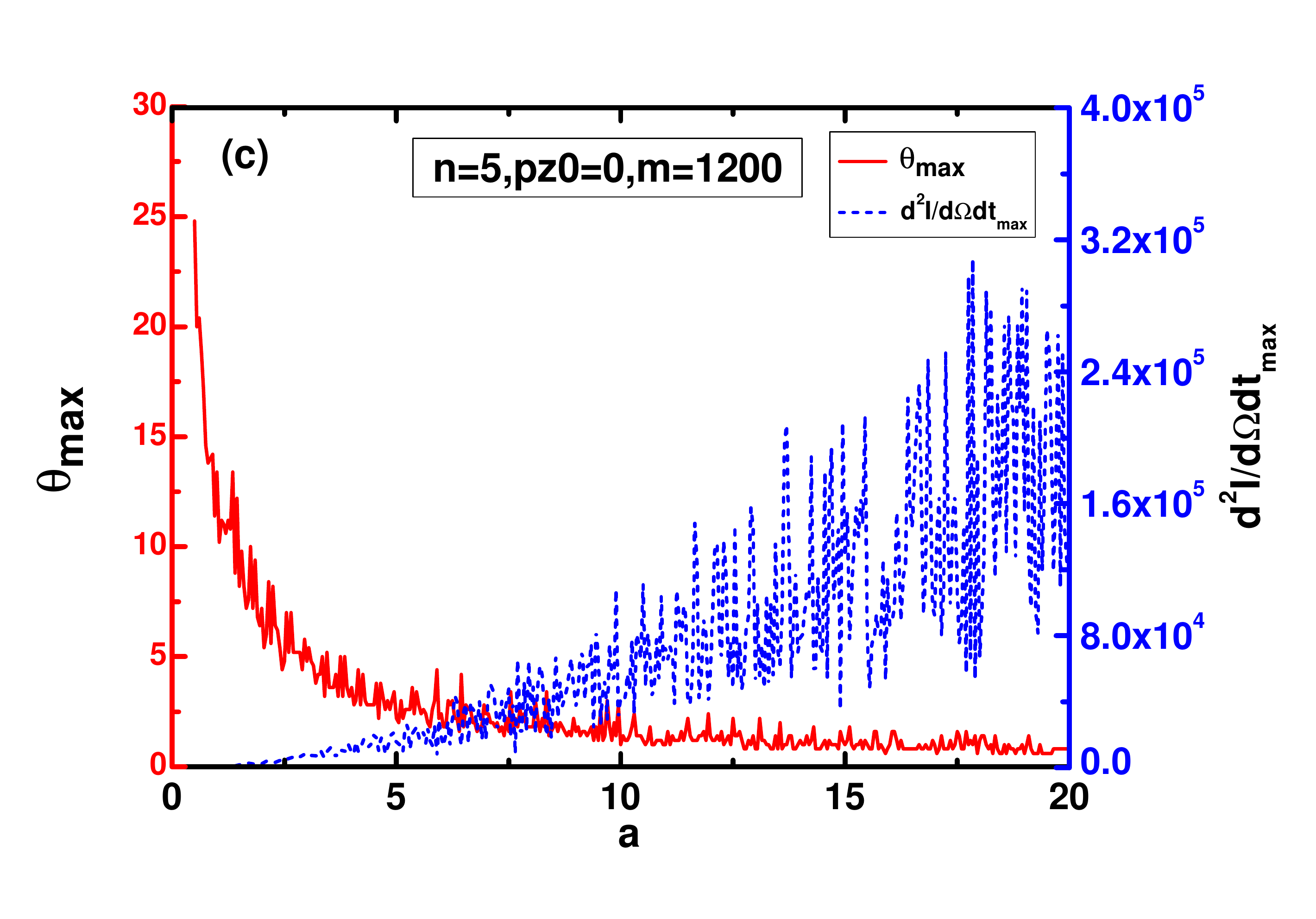}
\includegraphics[width=7.5cm]{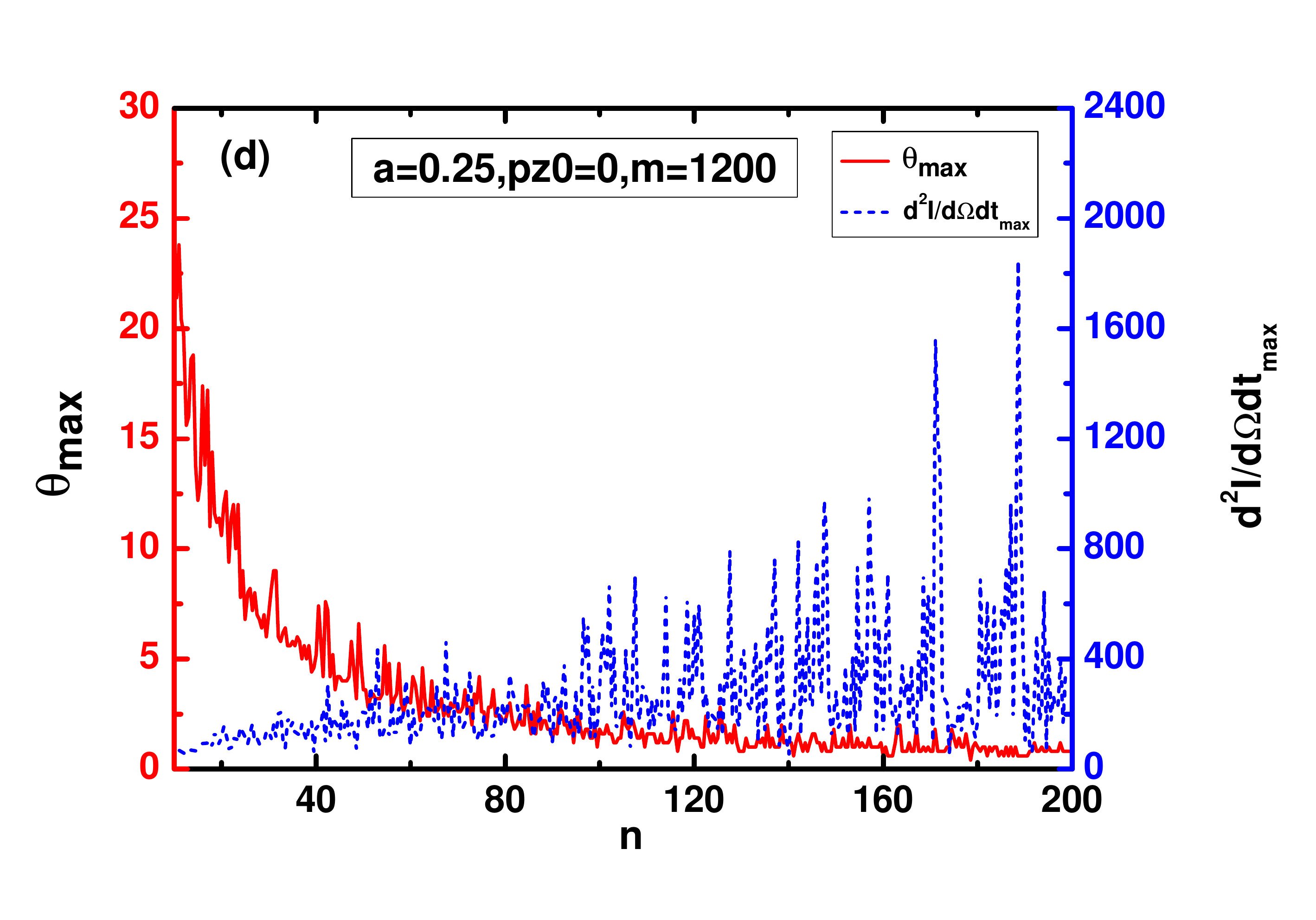}
\caption{\label{Fig5}(color online)The variation of the maximum radiation intensity  angle on various parameters for (a) the order of harmonics $m$, (b) the initial axial momentum $p_{z0}$, (c) the laser intensity $a$, (d) the resonance parameters $n$. The red solid line represents the angle and the blue short dash line represents the radiation intensity. The wavelength $\lambda=1\mu\textit{m}$, and the azimuthal angle $\varphi=0$. (a) $a=0.5$, $n=5$, $p_{z0}=0$. (b) $a=0.25$, $n=5$, $m=1200$. (c) $a=5$, $p_{z0}=0$, $m=1200$. (d) $a= 0.25$, $p_{z0}=0$, $m=1200$.}.
\end{figure}

From above, it is known that the radiation with respect to the polar angle $\theta$ is mainly distributed in two regions. In this section, we concentrate on one of them where the polar angle $\theta$ varies from 0 to $\pi$ to study the best observation angle $\theta_{max}$, that is the maximum radiation intensity angle, which is corresponding to the maximum radiation intensity $d^{2}I/d\Omega dt_{max}$. Fig.5 shows the evolution of the maximum radiation intensity angle $\theta_{max}$ with various parameters. For Fig.\ref{Fig5}(a), the parameters are chosen as the laser intensity $a=0.5$, the resonance parameter $n=5$, the initial axial momentum $p_{z0}=0$, and the order of harmonics $m$ varies from 0 to 5000. With the increase of harmonic order, it can be seen that the $\theta_{max}$ changes slightly between 17 and 27 degrees, which is consistent with the conclusion in the previous section. Another point showed from Fig.\ref{Fig5}(a) is that the radiation strength $d^{2}I/d\Omega dt_{max}$ decreases sharply to 0 when the harmonics order increases. when the harmonics order $m$ is between 1000 and 2000, the $\theta_{max}$ does not change much and the radiation strength $d^{2}I/d\Omega dt_{max}$ is probably moderate. So in the following three figures, the harmonics order $m=1200$ is fixed as an example. The variation of the $\theta_{max}$ and $d^{2}I/d\Omega dt_{max}$ with the initial axial momentum $p_{z0}$ is shown in Fig.\ref{Fig5}(b). To ignore the radiation reaction effect, the laser intensity $a=0.5$, the resonance parameter $n=5$ are chosen, and the initial axial momentum $p_{z0}$ varies from 0 to 20. It is apparent that the $\theta_{max}$ is getting smaller when the initial axial momentum $p_{z0}$ is increasing. When the initial axial momentum $p_{z0}$ is bigger than 10, we can see the $\theta_{max}$ is infinitely close to 0 degree. That is to say, if the electron moves fast enough along the $+z$ axis initially, the radiation can be focused on the forward. Influence of laser intensity on angular distribution of radiation is shown in Fig.\ref{Fig5}(c). It presents that the optimum observation direction $\theta_{max}$ changes observably when the laser's intensity is smaller than 6, and the radiation strengthens with the increase of the laser intensity. Besides, from Fig.\ref{Fig5}(c), It is also can be seen that when the laser intensity $a$ belongs to 6-20 interval, the radiation mainly distributes approximately within the range of 0-2.5 degrees from the $+z$ axis. For circularly polarized laser field, we have $a=\sqrt[2]{I_{0}\lambda_{\mu m}/(1.38\times2)}$. So, when $a=5, \lambda=1\mu m$, we can estimate the laser power density as $I_{0}=[{(1.38\times2)a^{2}}/{\lambda_{\mu m}^{2}}]\times10^{18}W/cm^{2}\approx7\times10^{19}W/cm^{2}$. Thus, it is worth noting that the RRE can be ignored when the laser field is weak (smaller than 6). Otherwise, the radiation of the electron is very complex and the RRE should be taken into consideration. Fig.\ref{Fig5}(d) shows the effect of the external magnetic field on the spatial distribution of radiation. One can see that with the increase of $n$, the $\theta_{max}$ gets smaller and the $d^{2}I/d\Omega dt_{max}$ increases. By the way, from our previous work Ref.\cite{EPL-117-44002}, the external magnetic field $B_{0}$ can be approximated as $B_{0}\approx(1+1/n)(100MG)$ when $p_{z0}\ll1$. Here, we choose $n=100,p_{z0}=0$, and the corresponding magnetic field $B_{0}\approx(1+1/100)(100MG)\approx100MG$. That is to say, when the intensity of the magnetic field reaches about 100 megabytes of Gauss, the radiation can be well observed from the forward direction($\theta_{max}\leq5$ degrees).

In addition, a very important result is found here. Taking
Fig.\ref{Fig5}(a) as an example, the laser intensity $a$ is 0.5,
the resonance parameter $n$ is 5, and the initial axial momentum
$p_{z0}$ equals 0. When the harmonics order is 1200,
$\theta_{max}$ is 25.2 degrees and then one can get the
radiation's frequency $\omega/\omega_{0}=1.5048\times10^{2}$. When
the harmonics order is 5000, $\theta_{max}$ is 22 degrees and the
according radiation's frequency
$\omega/\omega_{0}=1.1432\times10^{3}$. Thus, it is found that the
radiation¡¯s frequency $\omega$ is 100-1000 times  of
$\omega_{0}$, which indicates that the high frequency part of the
radiation can reach the frequency range of X-ray
($10^{17}Hz-10^{18}Hz$).

\section{conclusion}\label{Conclusion}

In this study, the angular distributions of Thomson scattering are researched in detail by a electron moving in combined laser and magnetic fields. Firstly, the results show that the electron experiences a helical-type periodic motion, which does not depend on the initial phase critically. Furthermore, the angular distribution with respect to the azimuthal angle $\phi$ shows interesting shape which is twofold symmetric. On the other hand, the spatial distribution of radiation with respect to the polar angle $\theta$ consists of two lobes collimated in the forward direction, which are roughly symmetric of the laser-propagating direction. The larger the resonance parameter $n$, the laser intensity $a$ and the initial axial momentum $p_{z0}$ are, the closer the radiation distributes to the laser-propagation direction. Accordingly, it is possible that radiation can be concentrated in the forward direction if an appropriate set of parameters are chosen. Finally, the best angle of observation $\theta_{max}$ expediently can be figured out and the high frequency part of the radiation can reach the range of X-ray. And at the best observation angles, the intensity of radiation can also be increased to $10-10^{8}$ orders of magnitude.

\section{acknowledgments}

Authors are grateful to the anonymous referees for helpful suggestions to improve manuscript. This work was supported by the National Natural Science Foundation of China (NSFC) under Grant Nos. 11475026 and 11305010. The computation was carried out at the HSCC of the Beijing Normal University.

\end{document}